\documentclass[superscriptaddress,secnumarabic,amssymb, nobibnotes, aps, prd]{revtex4-1}

\usepackage{amsmath}
\usepackage{graphicx}

\begin{document}

\title{Higher-Harmonic Collective Modes in a Trapped Gas from Second-Order Hydrodynamics}%

\author{W. E. Lewis}%
\email[Corresponding Author: ]{william.lewis-1@colorado.edu}
\affiliation{Department of Physics, University of Colorado, 390 UCB, Boulder, CO 80309, USA}

\author{P. Romatschke}
\affiliation{Department of Physics, University of Colorado, 390 UCB, Boulder, CO 80309, USA}
\affiliation{Center for Theory of Quantum Matter, University of Colorado, 390 UCB, Boulder, CO 80309, USA}
\date{August 29, 2016}%

\begin{abstract}
Utilizing a second-order hydrodynamics formalism, the dispersion relations for the frequencies and damping rates of collective oscillations as well as spatial structure of these modes up to the decapole oscillation in both two- and three- dimensional gas geometries are calculated. In addition to higher-order modes, 
the formalism also gives rise to purely damped ``non-hydrodynamic" modes. We calculate the amplitude of the various modes for both symmetric and asymmetric trap quenches, finding excellent agreement with an exact quantum mechanical calculation. We find that higher-order hydrodynamic modes are more sensitive to the value of shear viscosity, which may be of interest for the precision extraction of transport coefficients in Fermi gas systems.
\end{abstract}
\maketitle
%\tableofcontents

\section{Introduction}

Strongly interacting quantum fluids (SIQFs) such as high $T_c$ superconductors \cite{Rameau2014}, clean graphene \cite{Muller2009}, the quark-gluon plasma \cite{RHICWhite}, and Fermi gases tuned to a Feschbach resonance \cite{Book} seem to lack a description in terms of quasi-particle degrees of freedom. This has fueled interest in developing new tools to understand the transport properties of these fluids, as well as trying to experimentally determine those properties more precisely.

As of yet, one of the cleanest experimental realization of SIQFs is a Fermi gas tuned to a Feschbach resonance. Fermi gases offer unprecedented control of a multitude of properties such as interaction strength, system geometry, spin imbalance \cite{Liao2010,Ong2015}, and mass imbalance \cite{Trenkwalder2011}. In the case of a spin and mass balanced gas, there have been a number experiments aimed at the extraction of shear viscosity \cite{Schafer2007,Cao2011,Elliot2014,Joseph2015}, and --- to a lesser extent --- bulk viscosity \cite{CaoThesis}.

The Navier-Stokes equations provide a relatively straightforward model for the dependence of cloud expansion and collective oscillation phenomena on transport coefficients, making them a seemingly ideal candidate for extraction of such coefficients. Yet, in the low density corona of trapped atom gases the local mean free path becomes large, and hence one cannot expect the Navier-Stokes equations to apply. This, as well as uncertainties arising e.g. from trap averaging, gives rise to a large systematic error in transport coefficients thus extracted from experimental data. Hence a theory which can address both the hydrodynamic behavior of the high density region  as well as the low density corona of the cloud is desirable. It has been shown that by including extra ``non-hydrodynamic" degrees of freedom in a fluid dynamical description, termed anisotropic fluid dynamics, one can obtain a smooth crossover between Navier-Stokes dynamics in the high density core of the gas cloud and kinetic theory in the low density corona \cite{Bluhm2015}. This theory has been recently used to determine the shear viscosity in the high temperature regime with an error of five percent by comparing experimental data for an expanding cloud to an anisotropic hydrodynamic description \cite{Bluhm2016}. Similar precision determinations for transport properties at lower temperatures, e.g. close to the superfluid transition, are still outstanding.

This present work is related to studies using anisotropic hydrodynamics in the sense that we will also employ a hydrodynamic description beyond Navier-Stokes (``second-order hydrodynamics'') in order to study collective oscillations of harmonically trapped Fermi gases above the super-fluid transition ($T>T_c$). In the linear response regime we are considering in this work, it turns out that the second-order and anisotropic hydrodynamic equations of motion are identical. We will refer to our approach as second-order hydrodynamics to simplify the discussion, but the only difference to an anisotropic hydrodynamics framework will be in name.

Our work is closely related to Ref.\cite{Rosi2015}, and in many aspects is complementary to the results therein. In this work, the focus is on the effects arising from a non-vanishing shear viscosity, and we limit our consideration to ideal equation of state, whereas in Ref.~\cite{Rosi2015} collective modes for polytropic equations of state and zero shear viscosity were studied.

 The outline of the paper is as follows: We begin by describing our theoretical framework of second-order hydrodynamics in Sec.~\ref{SOH}. We then proceed to calculate the frequencies, damping rates, and spatial structures for the collective modes of harmonically trapped gases in both two and three dimensions in Secs.~\ref{2dmodes},\ref{3dmodes}. We calculate mode excitation amplitudes for experimentally relevant conditions in Sec.~\ref{ampcalcs} and offer our conclusions in Sec.~\ref{sec:conc}. Detailed results on the spatial mode structure and mode amplitude calculations can be found in three appendices.

\section{Second-Order Hydrodynamics}\label{SOH}

The Navier-Stokes equations are conservation equations for mass, momentum, and energy. To close the system of equations, constitutive relations between the viscous stresses and fluid variables need to supplied. For Navier-Stokes, the viscous stress tensor $\pi_{ij}$ is set to first-order gradients of the fluid dynamic variables, mass density $\rho$, flow velocity $\mathbf{u}$, and temperature $T$. While widely successful in many fluid dynamics applications, such a first order gradient expansion suffers from certain problems, in particular, in systems where the fluid speed approaches the speed of light \cite{Hiscock:1985zz}.  Thus, more recently a second-order hydrodynamic framework has been developed which -- true to its name -- includes second-order gradients in the expansion of the stress tensor, with appropriate new second-order transport coefficients. Unlike the similar framework of the Burnett equations, second-order hydrodynamics in addition contains a resummation procedure which ensures that it is a consistent, causal and instability-free generalization of Navier-Stokes (cf. the reviews in Refs.~\cite{Romatschke2010,Book}). For the case of a unitary Fermi gas, scale-invariance seems to be a good symmetry, and the resulting non-relativistic form of the second-order hydrodynamic equations has been derived in Ref.~\cite{Chao2012}. 

\subsection{Basic Equations}
 
In the following we consider what is maybe the simplest possible second-order hydrodynamics formalism to describe a fluid in $d$ spatial dimensions with a trapping force $\mathbf{F}$. Namely, we utilize a relaxation equation for the stress tensor.  In this case our fluid equations are given by
\begin{align}
	\label{masscons}
	&\partial_t\rho + \partial_i(\rho u_i)=0,\\
	\label{momcons}
	&\partial_t(\rho u_i)+ \partial_j(\rho u_i u_j + P \delta_{ij} + \pi_{ij})=\rho \frac{F_i}{m},\\
	\label{engcons}
	&\partial_t \epsilon + \partial_j\big[u_j\big(\epsilon + P\big)+\pi_{ij} u_i\big]=\rho \frac{F_k}{m} u_k,\\
	\label{stressten}
	 &\pi_{ij} + \tau_\pi \partial_t \pi_{ij} = -\eta \sigma_{ij},
\end{align}
where $\epsilon$ is the energy density, $\eta$ is the shear viscosity, and $\tau_\pi$ is the relaxation time for the stress tensor. In the above equations, $\epsilon$ and $\sigma_{ij}$ are specified in terms of the fluid velocity, mass density and pressure $P$ as 
\begin{align}
	\label{eng}
	&\epsilon=\frac{\big(\rho \mathbf{u}^2 + d P\big)}{2},\\
	\label{sig}
	&\sigma_{ij} = \big[\partial_i u_j + \partial_j u_i - \frac{2}{d} \delta_{ij} \partial_k u_k \big].
\end{align}
Note that Eq.~(\ref{eng}) corresponds to the equation of state for a scale-invariant system. It is easy to show that the familiar Navier-Stokes equations are recovered upon taking the limit $\tau_\pi\rightarrow 0$ in Eq.~(\ref{stressten}). 

\subsection{Assumptions}
For simplicity, we have assumed the bulk viscosity and heat conductivity coefficients to vanish. The assumption of vanishing bulk viscosity is consistent with measurements in two dimensions.\cite{TaylorPRL2012,VogtPRL2012}. Furthermore, calculations of bulk viscosity in $d=3$ imply that the value of bulk viscosity near unitarity in the high temperature limit should be small \cite{SchaferPRL2013}. Since we will consider a Fermi gas in the normal phase, \textit{i.e.} above the superfluid transition temperature $T_c$, taking the bulk viscosity to vanish should be a good approximation in the case $d=3$ as well. The assumption of vanishing thermal conductivity is justified as it is already a second-order gradient effect as discussed in Ref.~\cite{SchaferPRA2014}. Hence we assume the gas is isothermal, but it is straightforward to see how the procedure below can be extended to the non-isothermal case. As a consequence, the temperature is a function of time only and not of spatial coordinates. In order to obtain analytically tractable results, we additionally make the approximation that the gas may be described with an ideal equation of state:
\begin{equation}
	\label{eos}
	P = n T = \frac{\rho T}{m},
\end{equation}
where $n$ is the number density of particles (we let $\hbar=k_B=1$ throughout). The effects of a realistic non-ideal equation of state on collective mode behavior in a viscous fluid typically require numerical treatments such as those presented in Ref.~\cite{BrewerPRA2016}. 

Moreover, we assume $\eta/P$ to be constant. While this assumption is not expected to hold in the low density corona, it will allow analytic access to the spatial structure, frequency and damping rates of collective modes using a second-order hydrodynamics framework. More accurate numerical studies including temperature and density effects on the shear viscosity are left for future work. 

Finally, in order to access collective mode behavior of the gas, we will assume small perturbations around a time independent equilibrium state characterized by $\rho_0({\bf x})$, $\mathbf{u}_0=0$, and $T_{0}$ which are solutions to Eqs.~\eqref{masscons}-\eqref{eos}. Thus, we set $\rho = \rho_0 (1+ \delta \rho)$, $\mathbf{u}= \delta \mathbf{u}$, and $T= T_{0}+\delta T$ with $\delta \rho, \delta \mathbf{u}, \delta T$ assumed to be small. Working in the frequency domain we have $\delta \rho(t,\mathbf{x}) = e^{-i \omega t} \delta \rho(\mathbf{x})$, with similar expressions holding for $\delta \mathbf{u}$ and $\delta T$. To simplify notation, from now on perturbations such as $\delta \rho$ denote quantities where the time dependence has been factored out, unless otherwise stated.

\subsection{Linearization}

Expanding Eqs.~\eqref{masscons}-\eqref{stressten} to linear order in perturbations and utilize Eqs.~\eqref{eng}-\eqref{eos} assuming constant $\eta/P$, we have
\begin{align}
	\label{linmasscons}
	&-i \omega \rho_0 \delta \rho + \partial_i(\rho_0 \delta u_i)=0,\\
	\label{linmomcons}
	 &-i \omega \rho_0 \delta u_i + \partial_j(\frac{(T_0 \delta \rho + \delta T \rho_0)}{m} \delta_{ij} + \delta \pi_{ij})-\delta\rho \frac{F_i}{m}=0,\\
	\label{linengcons}	
	&-i \omega \frac{d}{2} \frac{(T_0 \delta \rho + \delta T \rho_0)}{m} + \partial_j \big( \delta u_j \frac{T_0 \rho_0}{m} \big)-\rho_0 \frac{F_k}{m} \delta u_k=0,\\
	\label{linstressten}	
	&\big(1-i \tau_\pi \omega\big) \delta \pi_{ij}=-\frac{\eta}{P} \frac{T_0 \rho_0}{m}  \big[\partial_i \delta u_j + \partial_j \delta u_i - \frac{2}{d} \delta_{ij} \partial_k \delta u_k \big].
\end{align}
We refer to Eqs.~\eqref{linmasscons}-\eqref{linengcons} with Eq.~\eqref{linstressten} substituted into Eq.~\eqref{linmomcons} as the linearized second-order hydrodynamics equations. It is straightforward to show that these equations  exactly match those arising from linearizing the anisotropic hydrodynamics framework of Ref.~\cite{Bluhm2015}.

\subsection{Configuration Space Expansion}\label{confexp}

For a harmonic trapping potential with trapping frequency $\omega_\perp$, the solution for the equilibrium density configuration is given by $\rho_0(\mathbf{x}) =\rho_0 \exp{[{\frac{-(x^2+y^2)m \omega_\perp^2}{2 T_0}}]}$. In the following, we will be using dimensionless units such that  all distances are measured in units of $(T_0/(m \omega_\perp^2))^{1/2}$, times are measured in units of $\omega_\perp^{-1}$, temperatures in units of $T_0$, and densities in units of $m^{d/2+1}\omega_\perp^d/T_0^{d/2}$. In these units the equilibrium solution is given by
\begin{align}
	\label{unitlessequil}
	\nonumber \rho_0(\mathbf{x}) =A_0 &e^{\frac{-(x^2+y^2)}{2 }}, \quad
	\mathbf{u}_0(\mathbf{x})=0,\quad T_0(\mathbf{x})= 1,
\end{align}
where $A_0$ is a dimensionless positive number setting the number of particles (cf. the discussion in App.~\ref{App3}).

In the absence of a trapping potential, it is usually convenient to perform a 
spatial Fourier transform of Eqs.~\eqref{linmasscons}-\eqref{linstressten} in order to obtain the collective modes of the system.  However, here we are interested in a harmonic trapping potential (linear trapping force) which breaks translation symmetry. Thus, it is more convenient to use a different expansion basis for the perturbations. Here we choose to expand perturbations in tensor Hermite polynomials, though any complete basis of linearly independent polynomials will do. The $N^{th}$ order tensor Hermite polynomials in $d$ spatial dimensions are given by the Rodrigues formula \cite{Coelho2014}
	\begin{equation}
		\label{Rodrigues}
		\text{H}^{\phantom{...}(N)}_{i_1 i_2 ... i_N}(\mathbf{x}) = \frac{(-1)^N}{g(\mathbf{x})}\frac{\partial^N g(\mathbf{x})}{\partial x^{i_1}\partial x^{i_2}...\partial x^{i_N}},\quad g(\mathbf{x}) = \frac{1}{(2 \pi)^{\frac{d}{2}}}e^{\frac{-\mathbf{x}^2}{2}},
	\end{equation}
where $i_k \in \{1,2,...d\}$ for $k=\{1,2,...,N\}$.
The tensor Hermite polynomials are orthogonal with respect to a Gaussian weight which makes them particularly useful for the case of a harmonic trapping potential. In particular, they satisfy the orthonormality condition
	\begin{equation}
		\int d^d\mathbf{x} \phantom{.}g(\mathbf{x}) \phantom{.}\text{H}^{\phantom{...}(N)}_{i_1 i_2 ... i_N}(\mathbf{x}) \phantom{.}\text{H}^{\phantom{...}(M)}_{j_1 j_2 ... j_M}(\mathbf{x}) = \delta^{NM} (\delta^{i_1j_1}\delta^{i_2j_2}...\delta^{i_Nj_N}+\text{all permutations of $j$'s}).
	\end{equation}
Assuming translational invariance along the z-axis in $d=3$ spatial dimensions, the expansion in both $d=2$ and $d=3$ will involve only the tensor Hermite polynomials for $d=2$. Recalling the assumption that the gas is isothermal, the polynomial expansion of perturbations is then given by
\begin{align}
	\label{perts}
	\nonumber &\delta \rho(\mathbf{x}) =\sum_{M=0}^N \sum_{j=1}^{M+1} a^{(M)}_{\mathbf{m}_j}\text{H}^{(M)}_{\mathbf{m}_j}(\mathbf{x}),\\
	&\delta \mathbf{u}(\mathbf{x})=\sum_{M=0}^N \sum_{j=1}^{M+1} \mathbf{b}^{(M)}_{\mathbf{m}_j}\text{H}^{(M)}_{\mathbf{m}_j}(\mathbf{x}),\\
	\nonumber &\delta T(\mathbf{x})= c,
\end{align}
where, in the sum over $j$, $\mathbf{m}_j$ is understood to run over all combinations of indices unique up to permutations. For example, if $M=2$ the second sum runs over $\mathbf{m}=\{(1,1),(1,2),(2,2)\}$, while $(2,1)$ is excluded. The reason for this restriction is that $\text{H}^{(M)}_{\mathbf{m}}(\mathbf{x})$ is fully symmetric in the indices as can be seen from Eq.~\eqref{Rodrigues}. One should also note that in Eqs.~\eqref{perts}, $\mathbf{b}^{(M)}_{\mathbf{m}}$ is used as shorthand for the polynomial coefficients of all components of $\delta \mathbf{u}$, and for a given $M$ and $\mathbf{m}$ is a column vector with $d$ components. 

Let us now discuss the details of accessing the collective modes whose spatial structure is associated with polynomials of low degree (``low-lying modes''). Substituting Eqs.~\eqref{perts} truncated at polynomial order $N$ into the linearized second-order hydrodynamics equations and taking projections onto different tensor Hermite polynomials of order $K\leq N$ we obtain a matrix equation for the polynomial coefficients in Eqs.~\eqref{perts}. The (complex) collective mode frequencies $\tilde \omega$ are then obtained from requiring a non-trivial null-space of this matrix, and subsequently the spatial structures are obtained from the corresponding null-vectors.

\section{Collective Mode Solutions in $d=2$}\label{2dmodes}

\begin{figure*}[ht!]
\includegraphics[width=\textwidth]{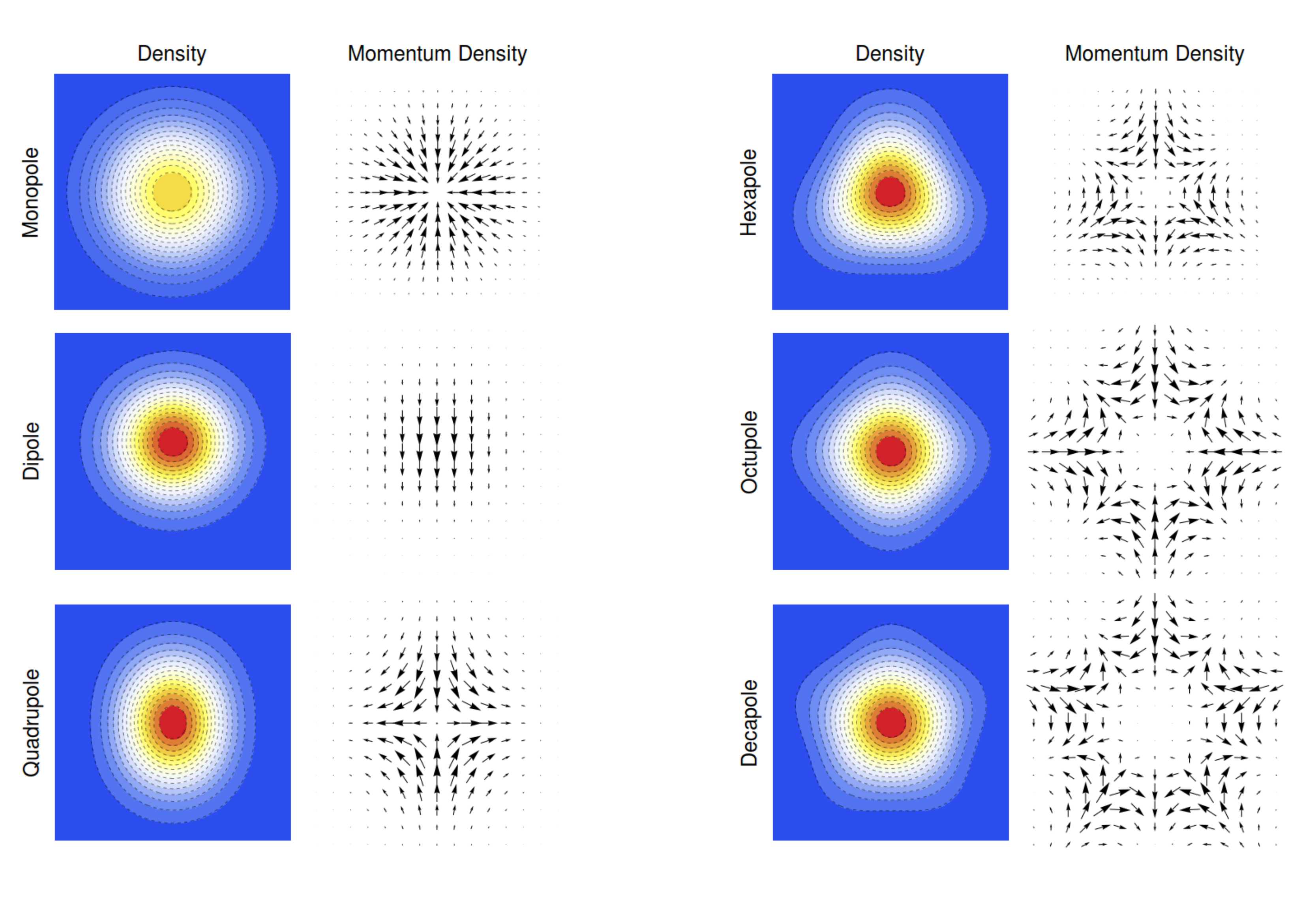}
  \caption{Time snap shots of density profiles and subsequent momentum density ($\rho\mathbf{u}$) for the oscillatory modes in $d=2$. Note that the center of the monopole mode is at a lower density than the centers of the other modes since it is volume changing and has a larger radius than the equilibrium configuration. The damping rate of higher-order modes is more sensitive to $\eta/P$ as discussed in the text. Also note that non-hydrodynamic modes share the same spatial structure as their hydrodynamic counterpart.}
  \label{fig:key}
\end{figure*}

Results for the density and velocity of low-lying collective modes in $d=2$ are shown in Fig.~\ref{fig:key}. In particular, we find a breathing (monopole) mode which corresponds to a cylindrically symmetric oscillatory change in cloud volume, a sloshing (or dipole) mode where the center of mass of the cloud oscillates about the trap center, a quadrupole mode which is elliptical in shape, and higher-order modes corresponding to higher-order geometric shapes. Note that the spatial structure of these collective modes are similar to those reported in Ref.~\cite{Rosi2015}. 
More detailed information about the $d=2$ collective modes can be found in App.~\ref{App1}.

\begin{table}[h!]
\begin{center}
\begin{tabular}{| c | c | c |}
  \hline
  \phantom{...}&$\omega$ & $\Gamma$\\
  \hline			
  Number (Zero Mode) & $0$ & $0$ \\
  \hline			
  Temperature (Zero Mode) & $0$ & $0$\\
  \hline			
  Rotation (Zero Mode) & $0$ & $0$\\
  \hline			
  Breathing (Monopole) & 2 & 0 \\
  \hline
  Sloshing (Dipole) & 1 & 0 \\
  \hline
  Quadrupole & $\sqrt{2}$ & $\frac{\eta}{P}$ \\
  \hline
  Hexapole & $\sqrt{3}$ & $2\frac{\eta}{P}$  \\
  \hline
  Octupole & 2 & $3\frac{\eta}{P}$  \\
  \hline
  Decapole & $\sqrt{5}$ & $4\frac{\eta}{P}$  \\
  \hline  
  Non-hydrodynamic Quadrupole& 0 & $\frac{1}{\tau_\pi} - 2\frac{\eta}{P}$ \\
    \hline  
  Non-hydrodynamic Hexapole & 0 & $\frac{1}{\tau_\pi } - 4\frac{\eta}{P}$ \\
    \hline  
  Non-hydrodynamic Octupole & 0 & $\frac{1}{\tau_\pi } - 6\frac{\eta}{P}$ \\
  \hline  
  Non-hydrodynamic Decapole & 0 & $\frac{1}{\tau_\pi } - 8\frac{\eta}{P}$ \\
  \hline
\end{tabular}
\end{center}
  \caption{Frequencies and damping rates in $d=2$ from linearized second-order hydrodynamics assuming $\frac{\eta}{P},\tau_\pi\ll 1$. The hydrodynamic mode damping rates depend on $\eta/P$ times a prefactor which increases with mode order. Note that for $d=2$ there is no non-hydrodynamic sloshing or breathing mode.  }
\label{2dfreqtab}
\end{table}

The collective mode frequencies $\omega$ and damping rates $\Gamma$ are given as the real and imaginary parts of roots of polynomials, which generally do not admit simple closed form expressions. Hence, in Tab.~\ref{2dfreqtab} we choose to report expressions for the complex frequencies and spatial mode structure from second-order hydrodynamics for the low-lying modes in the hydrodynamic limit $\eta/P \ll 1$ and $\tau_\pi \ll 1$ (assuming that $\tau_\pi$ and $\eta/P$ are of the same order of magnitude), in which case simple analytic expressions can be obtained. In addition to the modes shown in Fig.~\ref{fig:key} there are three modes in Tab.~\ref{2dfreqtab} which have zero complex frequency. The first corresponds to a change in total particle number, the second corresponds to a change in temperature and width of the cloud, and the third ``zero mode" is simply a rotation of the fluid about the central axis. While they are required for the mode amplitude analysis (see Sec.~\ref{ampcalcs}), the role of the first two of these zero frequency modes is relatively uninteresting. Hence, we relegate detailed discussion of these modes to App.~\ref{App3}.

\begin{figure*}
  \includegraphics[width=\textwidth]{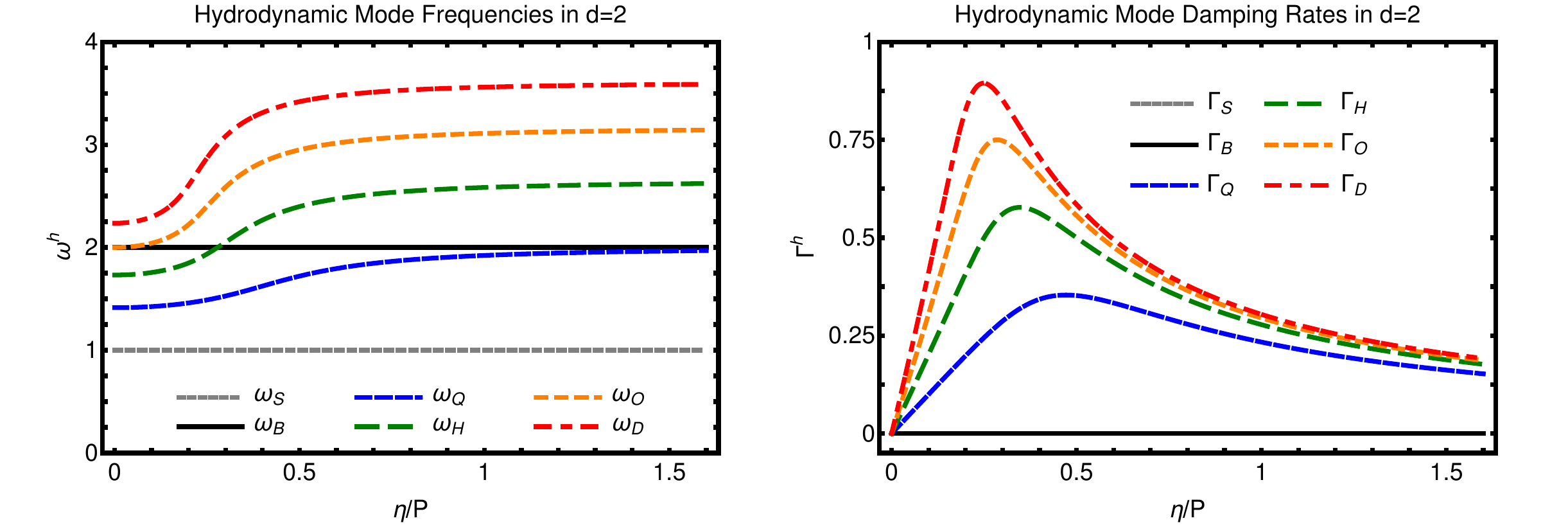}%
  \caption{Two-dimensional hydrodynamic collective mode frequencies $\omega$ (left panel) and damping rates $\Gamma$ (right panel) as a function of $\eta/P$. Subscripts denote mode name (monopole ``B''; dipole ``S''; quadrupole ``Q''; hexapole ``H''; octupole ``O''; decapole ``D''). For the purpose of the figures the kinetic theory relation $\tau_\pi=\eta/P$ has been used.
}
  \label{fig:2dfreq}
\end{figure*}

The rows of Tab.~\ref{2dfreqtab} starting with the number mode and ending with the decapole mode are all hydrodynamic modes. We note that at order $\mathcal{O}(\eta/P)$ the results for these modes match those from an analysis of the mode frequencies of the Navier-Stokes equations at the same order. However, for values of $\eta/P$ where corrections to the hydrodynamic limit become significant, the frequencies found from the Navier-Stokes equations and second-order hydrodynamics disagree. In Fig.~\ref{fig:2dfreq} we show the full dependence of the hydrodynamic mode frequencies and damping rates on $\eta/P$ (assuming $\tau_\pi=\eta/P$ based on kinetic theory \cite{Bruun2007,SchaferPRA2014,Kikuchi2016}). Note that the result of second-order hydrodynamics for the quadrupole mode exactly matches the result from kinetic theory when setting $\tau_\pi=\tau_R=\eta/P$ \cite{Baur2013,BrewerPRA2016}.

Furthermore,  results shown in in Tab.~\ref{2dfreqtab} demonstrate that the hydrodynamic mode damping rates depend on $\eta/P$ times a prefactor which increases with mode order. This is completely analogous to what has been observed in experiments on relativistic ion collisions, where simultaneous measurements of multiple modes have been used to obtain strong constraints on the value of $\eta/s$, cf. Ref.~\cite{Schenke2012}. While higher-order modes have not yet been studied in experiment, it is conceivable that measuring their damping rates could lead to a similarly strong experimental constraint on shear viscosity in the unitary Fermi gas. We are not aware of this approach having been suggested elsewhere in the literature. When aiming for using higher-order modes to analyze shear viscosity in Fermi gases we recall that the present analysis is based on a linear response treatment. Quantitative analysis of higher-order flows will, however, require the inclusion of nonlinear effects, especially for analysis of flows beyond hexapolar order due to mode mixing. For this reason, we suggest the hexapolar mode as a prime candidate for the use of higher-order modes to extract shear viscosity. 

\begin{figure}[htbp]
      \includegraphics[width=0.5\textwidth]{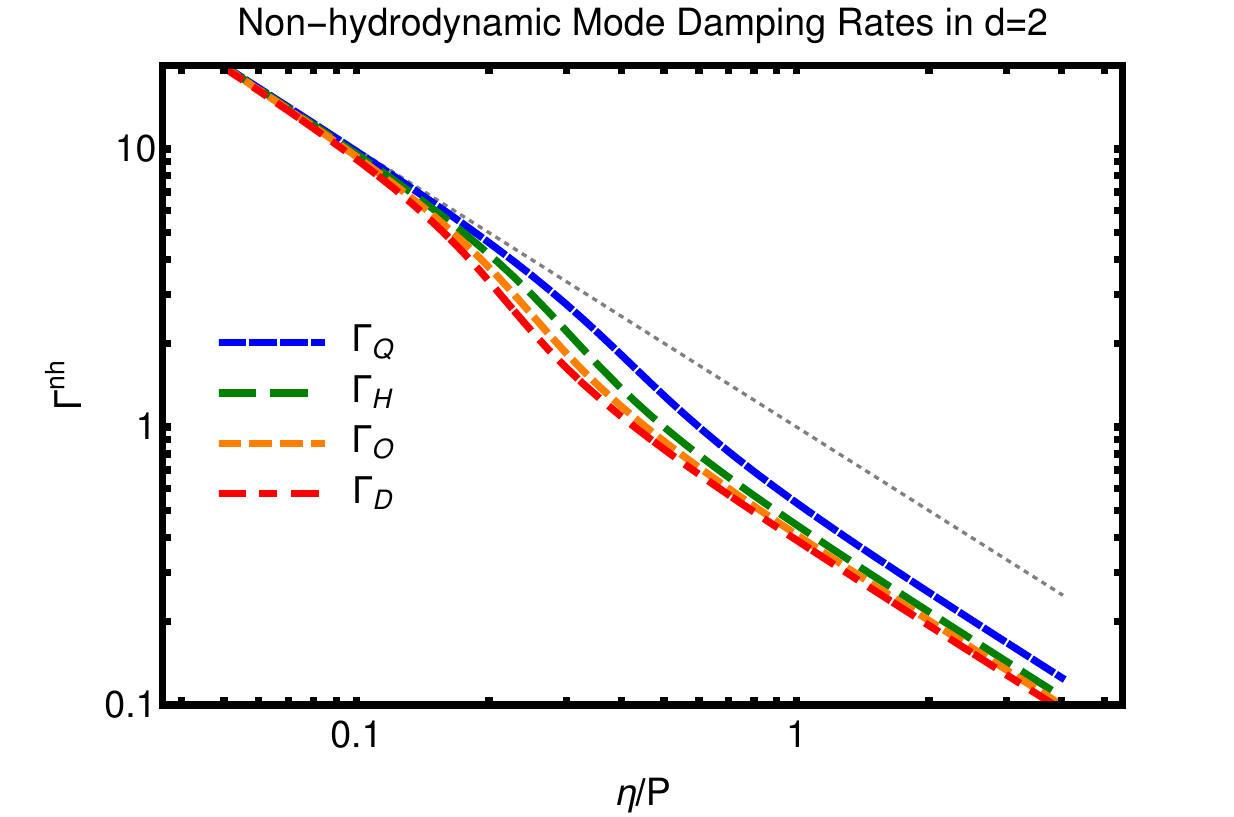}
          \caption{Two-dimensional non-hydrodynamic collective mode damping rates $\Gamma$ as a function of $\eta/P$ (using $\tau_\pi=\eta/P$). Subscripts denote mode name (quadrupole ``Q''; hexapole ``H''; octupole ``O''; decapole ``D''). Dotted line is $\Gamma^{nh}= 1/\tau_\pi$ (cf. Ref.~\cite{BrewerPRL2015}). Note that in $d=2$ there is no non-hydrodynamic sloshing or breathing mode.}
      \label{2dnhdamp}
\end{figure}

 Finally, Tab.~\ref{2dfreqtab} also indicates the presence of non-hydrodynamic modes (e.g. modes not present in a Navier-Stokes description). The physics of non-hydrodynamic modes is largely unexplored (cf. Refs.~\cite{BrewerPRL2015,Bantilan:2016qos} for a brief discussion of the topic in the context of cold quantum gases). Results shown in Tab.~\ref{2dfreqtab} imply that several such non-hydrodynamic modes exist, all of which are purely damped in second-order hydrodynamics. The non-hydrodynamic mode damping rates are sensitive to $\tau_{\pi}$ and $\eta/P$. Thus the value of $\tau_\pi$ could be extracted experimentally by measuring any of the non-hydrodynamic mode damping rates in combination with a hydrodynamic mode damping rate required to determine $\frac{\eta}{P}$. In Fig.~\ref{2dnhdamp}, non-hydrodynamic damping rates are shown as a function of $\eta/P$ when setting $\tau_\pi=\eta/P$.

\section{Collective Mode Solutions in $d=3$}\label{3dmodes}

In the case of a three-dimensional gas in a harmonic trap with trapping frequencies $\omega_z\ll\omega_x=\omega_y$, the resulting gas cloud takes on an elongated cigar shaped geometry. For $\omega_z=0$, the configuration space expansion Eqs.~\eqref{perts} can be applied because there is no dependence on the coordinate $z$ if we assume a translationally invariant system along the $z$-axis. In this case, the collective mode structures in $d=3$ are qualitatively similar to those obtained in the two-dimensional case, cf. the discussion in Refs.~\cite{KinastPRL2005,AltmeyerPRA2007}.

\begin{figure*}[htbp]
      \includegraphics[width=1\textwidth]{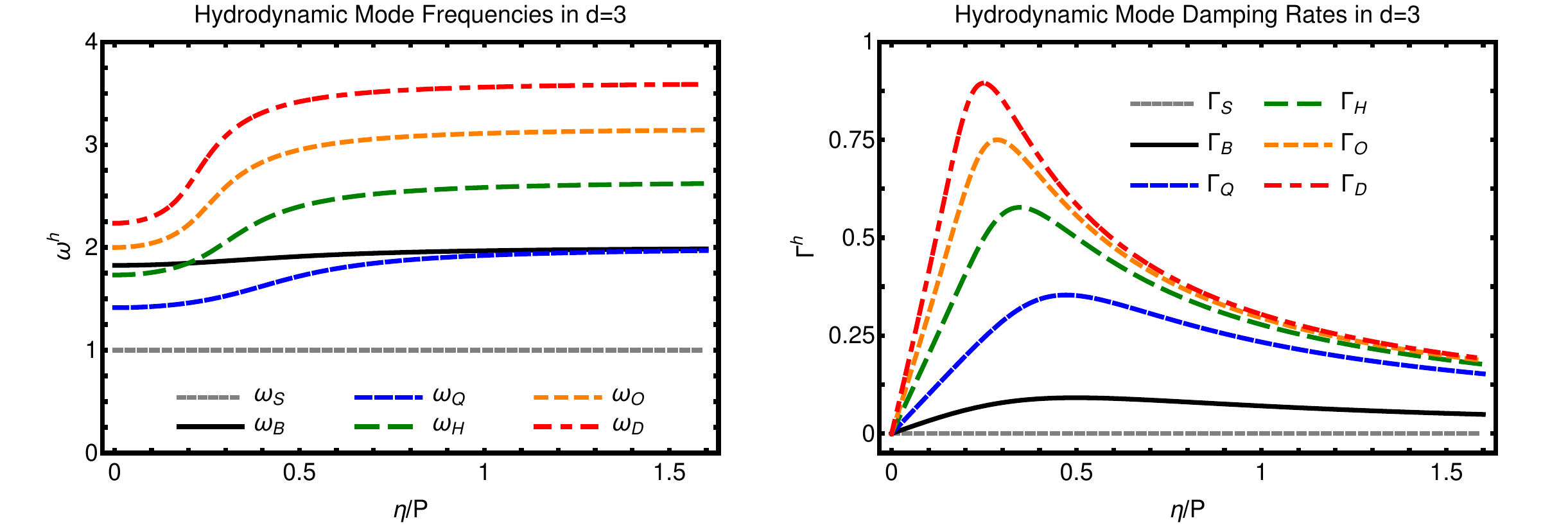}
      \caption{Three-dimensional hydrodynamic collective mode frequencies $\omega$ (left panel) and damping rates $\Gamma$ (right panel) as a function of $\eta/P$. Subscripts denote mode name (monopole ``B''; dipole ``S''; quadrupole ``Q''; hexapole ``H''; octupole ``O''; decapole ``D''). For the purpose of the figures the kinetic theory relation $\tau_\pi=\eta/P$ has been used.}
      \label{3dfreq}
\end{figure*}

\begin{figure}[htbp]
     \includegraphics[width=0.5\textwidth]{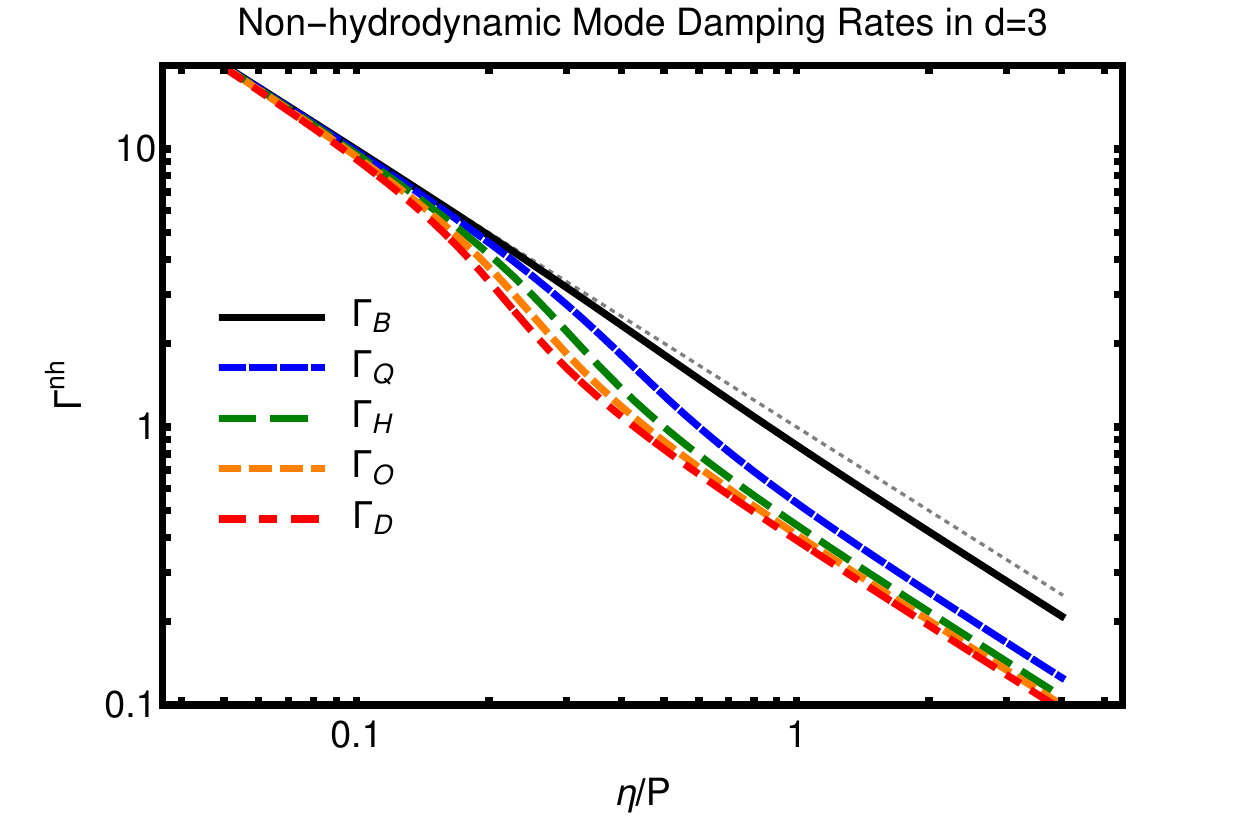}
    \caption{Three-dimensional non-hydrodynamic collective mode damping rates $\Gamma$ as a function of $\eta/P$ (using $\tau_\pi=\eta/P$). Subscripts denote mode name (monopole ``B''; quadrupole ``Q''; hexapole ``H''; octupole ``O''; decapole ``D''). Dotted line is $\Gamma^{nh}= 1/\tau_\pi$ (cf. Ref.~\cite{BrewerPRL2015}).  Note that in $d=2$ there is no non-hydrodynamic sloshing mode, but there is a non-hydrodynamic breathing mode.}
      \label{3dnhdamp}
\end{figure}

We report results for the low-lying modes in the limit $\eta/P,\tau_\pi\ll 1$ in Tab.~\ref{3dfreqtab} whereas the full dependence of frequencies and damping rates on $\frac{\eta}{P}$ is shown in Figs.~\ref{3dfreq},\ref{3dnhdamp} for the case  $\tau_\pi=\eta/P$.
The only qualitative difference with respect to the $d=2$ case is that the breathing mode in $d=3$ has a different frequency, a non-zero damping rate, and there is now a non-hydrodynamic breathing mode. See App.~\ref{App2} for more details about the spatial structure of the d=3 collective modes.
\begin{table}
\begin{center}
\begin{tabular}{| c | c | c |}
  \hline
  \phantom{...}&$\omega$ & $\Gamma$\\
  \hline			
  Temperature (Zero Mode) & $0$ & $0$\\
  \hline			
  Number (Zero Mode) & $0$ & $0$ \\
  \hline			
  Rotation (Zero Mode) & $0$ & $0$\\
  \hline			
  Breathing (Monopole) & $\sqrt{\frac{10}{3}}$  & $\frac{\eta}{3P}$ \\
  \hline
  Sloshing (Dipole) & 1 & 0 \\
  \hline
  Quadrupole & $\sqrt{2}$ & $\frac{\eta}{P}$ \\
  \hline
  Hexapole & $\sqrt{3}$ & $2\frac{\eta}{P}$  \\
  \hline
  Octupole & 2 & $3\frac{\eta}{P}$  \\
  \hline
  Decapole & $\sqrt{5}$ & $4\frac{\eta}{P}$  \\
   \hline  
  Non-hydrodynamic Breathing& 0 & $\frac{1}{\tau_\pi} - \frac{2\eta}{3P}$ \\
  \hline  
  Non-hydrodynamic Quadrupole& 0 & $\frac{1}{\tau_\pi} - 2\frac{\eta}{P}$ \\
    \hline  
  Non-hydrodynamic Hexapole & 0 & $\frac{1}{\tau_\pi} - 4\frac{\eta}{P}$ \\
    \hline  
  Non-hydrodynamic Octupole & 0 & $\frac{1}{\tau_\pi} - 6\frac{\eta}{P}$ \\
  \hline  
  Non-hydrodynamic Decapole & 0 & $\frac{1}{\tau_\pi} - 8\frac{\eta}{P}$ \\
  \hline
\end{tabular}
\end{center}
  \caption{Frequencies and damping rates in $d=3$ from linearized second-order hydrodynamics assuming $\frac{\eta}{P},\tau_\pi\ll 1$. The hydrodynamic mode damping rates depend on $\eta/P$ times a prefactor with increases with mode order. Note that there is no non-hydrodynamic sloshing mode, but, unlike for $d=2$, there is a non-hydrodynamic breathing mode for $d=3$.}
\label{3dfreqtab}
\end{table}

It should be pointed out that, while second-order hydrodynamics predicts purely damped non-hydrodynamic modes for both $d=2,3$, more general (string-theory-based) calculations suggest that there should be a non-vanishing frequency component in the case of $d=3$ \cite{Bantilan:2016qos}. It would be interesting to measure non-hydrodynamic mode frequencies and damping rates in order to describe transport beyond Navier-Stokes on a quantitative level.

\section{Mode Amplitudes Calculations}\label{ampcalcs}

In this section, experimentally relevant scenarios to excite the collective modes of the previous sections are discussed, and the corresponding mode amplitudes are calculated. For simplicity, we assume $\tau_\pi=\eta/P$ in the following. In particular, we focus on studying the excitation of the non-hydrodynamic quadrupole (in $d=2,3$) and non-hydrodynamic breathing (in $d=3$) modes, leaving a study of higher-order modes for future work. For simplicity, only simple trap quenches (rapid changes in trap configuration) are considered. We will assume the gas cloud to start in an equilibrium configuration of a (possibly biaxial, i.e. $\omega_{x,\, init} \neq \omega_{y,\, init}$) harmonic trap. At some initial time, a rapid quench will bring the trap configuration into a final harmonic form, which is assumed to be isotropic in the x-y plane with trapping frequency $\omega_{x,\, final}=\omega_{y,\, final}=1$ in our units. 

In the case of Navier-Stokes equations, initial conditions are fully specified through the initial density $\rho_{\,init}$, velocity $\mathbf{u}_{init}$, and temperature $T_{\,init}$ or appropriate time derivatives of such quantities. However,  second-order hydrodynamics treats the stress tensor $\pi_{ij}$ as a hydrodynamic variable, so, in addition, an initial condition $\pi_{ij,\, init}$ or its time derivative needs to be specified.

For equilibrium initial conditions of a general biaxial harmonic trap with trapping force given by $\mathbf{F} = -\gamma_x \mathbf{x} - \gamma_y \mathbf{y}$   we have
	\begin{align}
		\rho_{init}(\mathbf{x}) = A_{i} &e^{\frac{-(\sigma_x x^2 + \sigma_y y^2)}{2 T_{\, init}}},\\
		\mathbf{u}_{init}&=0,\\
		\pi_{ij,\, init}&=0,
	\end{align} 
where $T_{\,init}$ also needs to be specified. Initial equilibrium implies the condition $\gamma_i = \omega_{i,\, init}^2 = \sigma_i /T_{\, init}$ so that the cloud width is fully specified once $\gamma_i$ for $i=x,y$ and $T_{\, init}$ are fixed. In addition, equilibrium of the initial trap allows us to take $\pi_{ij,\, init}=0$. The mode amplitudes can then be obtained by projecting initial conditions onto the collective modes found in the preceding sections (see App.~\ref{App3} for details of the calculation).

\paragraph*{Isotropic Trap Quench in $d=2$}\label{2diso}

\begin{figure*}[htbp]
    \begin{center}
      \includegraphics[width=\textwidth]{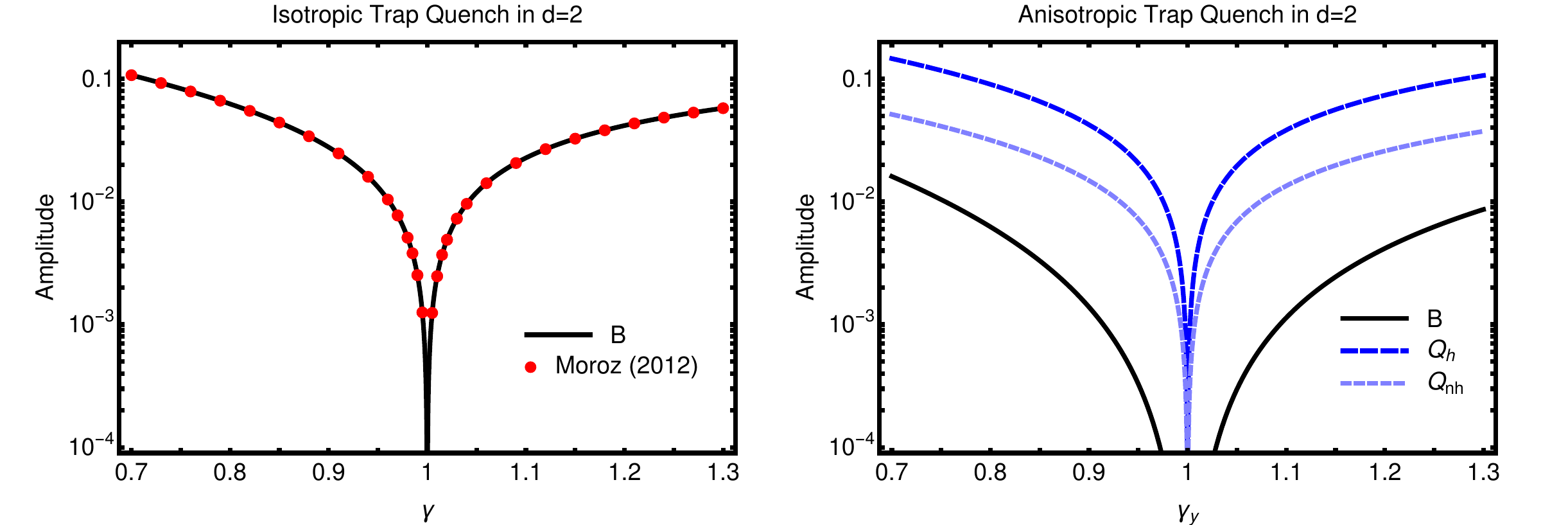}
      \caption{Left: Absolute value of the (dimensionless) breathing (``B") mode amplitude as a function the quench strength parameter $\gamma$ for an isotropic trap quench in d=2. Results are independent of $\frac{\eta}{P}$. The mode amplitude agrees perfectly with the result from an exact quantum mechanical scaling solution (``Moroz 2012") derived in Ref.~\cite{Sergej2012}. Right: Absolute value of the (dimensionless) breathing (``B"),  hydrodynamic (``Q$_h$") and non-hydrodynamic (``Q$_{nh}$") quadrupole mode amplitudes as a function the quench strength parameter $\gamma_y$ for an anisotropic trap quench in d=2. Results shown are for $\frac{\eta}{P}=0.5$. Note that the temperature mode amplitude (not shown) matches the breathing mode amplitude for both the isotropic and anisotropic trap quench in d=2.}
      \label{fig:2disoamp}
    \end{center}
\end{figure*}

We first consider the case of an isotropic trap quench $\gamma_x=\gamma_y\equiv \gamma$ in $d=2$ and assume $A_{i}/A_0=1$ and $T_{\, init}=1$ for simplicity. Although this case does not exhibit non-hydrodynamic or higher-order collective mode excitation, it does allow us to make direct comparison to results from the literature for the breathing mode excitation amplitude. This type of initial condition corresponds to a rotationally symmetric trap quench with no initial fluid angular momentum. Symmetry then implies that only the number, temperature, and breathing modes can be excited (cf. Tab.~\ref{2dfreqtab}), and the initial amplitude for these modes are readily calculated. Fig.~\ref{fig:2disoamp} displays the (dimensionless) breathing mode amplitude as a function of the quench strength $\gamma$. (Note that the amplitude of the temperature mode is identical to the breathing mode amplitude in this case.) The number mode is not excited since the number of atoms taken in the initial condition match the number of atoms we assumed in our final trap equilibrium ($A_{i}/A_0=1$).
The amplitude of the breathing mode for the isotropic trap quench is compared to the results from an exact quantum mechanical scaling solution by Moroz \cite{Sergej2012} in Fig.~\ref{fig:2disoamp}. As can be seen from this figure, there is exact agreement between the calculations for all strength values $\gamma$.
Note that the amplitudes in this case are independent of $\eta/P$ since for $d=2$, the breathing mode does not couple to the shear stress tensor $\pi_{ij}$.

\paragraph*{Anisotropic Trap Quench in $d=2$}

We perform a similar analysis to that above, considering the case $A_i/A_0=1$, and $T_{\, init}=1$, but now taking $\gamma_x\gamma_y=1$, which corresponds to an anisotropic trap quench. The mode amplitudes in this case depend on the value of $\eta/P$.  In this case, in particular the temperature, breathing and quadrupole modes are excited. Fig.~\ref{fig:2disoamp} shows the absolute value of the mode amplitudes for the hydrodynamic breathing and quadrupole modes, as well as the non-hydrodynamic quadrupole mode as a function of the quench strength $\gamma_y$. Not surprisingly, Fig.~\ref{fig:2disoamp} shows that the anisotropic trap quench gives rise to a considerably larger quadrupole mode amplitude (both hydrodynamic and non-hydrodynamic) than the amplitude of the breathing mode.

\begin{figure}[htbp]
      \includegraphics[width=\textwidth]{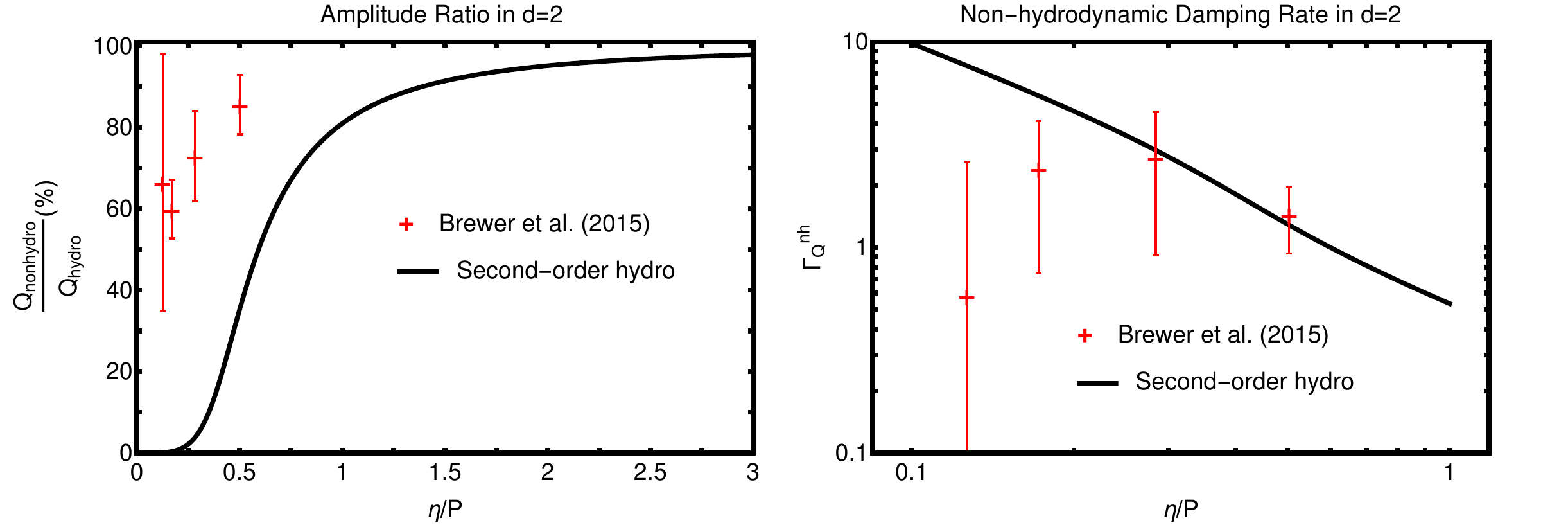}
      \caption{Comparison of second-order hydro theory (lines) and experimental (``Brewer et al. 2015") results as a function of $\eta/P$ for an anisotropic trap quench (results independent of $\gamma$). Left: ratio of non-hydro to hydro quadrupole mode amplitudes. Right:  non-hydrodynamic quadrupole mode damping rate $\Gamma$. Experimental data is from the reanalysis done in Ref.~\cite{BrewerPRL2015}.}
      \label{fig:anisovseta}
\end{figure}

For a potential experimental observation of the non-hydrodynamic quadrupole mode, it is interesting to consider the relative amplitude of this non-hydrodynamic mode to the (readily observable) hydrodynamic quadrupole mode. The (absolute) amplitude ratio calculated using the above anisotropic trap quench initial condition is plotted in Fig.~\ref{fig:anisovseta} as a function of $\eta/P$. One finds that the non-hydrodynamic mode amplitude is monotonically increasing as a function of $\eta/P$. This is plausible given that for small viscosities one expects the hydrodynamic mode to be dominant, whereas one expects the non-hydrodynamic mode to dominate in the ballistic $\eta/P\rightarrow \infty$ limit. 

The present calculation is compared to mode amplitude ratios extracted from experimental data \cite{VogtPRL2012} in Ref.~\cite{BrewerPRL2015}. To compare non-hydrodynamic damping rate data and theory, we follow the procedure used in Ref.~\cite{BrewerPRL2015} by employing the approximate kinetic theory relation
\begin{equation}
	\label{pauletatokfa}
 \frac{\eta}{P} \approx K[1+\frac{4 \ln^2(k_Fa)}{\pi^2}],
\end{equation}
where $K\approx 0.12$ in order to relate the experimentally determined $k_F a$ to $\eta/P$ (see discussion in Refs.~\cite{BrewerPRL2015,BrewerPRA2016} and references therein for more details on this relation). Using this procedure, one observes qualitative agreement of the amplitude ratios between calculation and experimental data in Fig.~\ref{fig:anisovseta} (left panel). In addition, one can compare the non-hydrodynamic quadrupole mode damping rate, finding reasonable agreement (cf. right panel of Fig.~\ref{fig:anisovseta}).

There are several possible reasons why a quantitative agreement between second-order hydro and experiment in Fig.~\ref{fig:anisovseta} should not be expected. For instance, the present theory calculations neglect the presence of a pseudogap phase
and pairing correlations (see e.g. Refs.~\cite{Baur2013,EnssPRL2014,Feld2011,Guo2011,SchaferPRA2012} on this topic). Furthermore, it is likely that the quantitative disagreement in Fig.~\ref{fig:anisovseta} is at least in part due to the assumptions discussed in Sec.~\ref{SOH}, such as small perturbations, constant $\eta/P$ and ideal equation of state. In particular, for the strongly interacting quasi-two-dimensional Fermi gas near the pseudogap temperature $T^*$, significant modification of the equation of state have been predicted and observed, cf. Ref.~\cite{EnssPRL2014}. Studies aiming for achieving a quantitative agreement most likely will have to rely on full numerical solutions, such as e.g. those discussed in Refs.~\cite{BrewerPRA2016,Bluhm2016}, which we leave for future work. In addition, our framework only admits a single non-hydrodynamic mode for each of the collective modes. While this may be appropriate in the kinetic theory regime, other approaches such as that of Ref. \cite{Bantilan:2016qos} indicate that our model may be too simple to capture quantitative features of early time dynamics. Finally, we note that the data shown in Fig.~\ref{fig:anisovseta} was extracted from experiments that were not designed with the purpose of considering early time dynamics. This may contribute to the large uncertainty of the existing data, as well as possibly introducing significant systematic error.

\paragraph*{Isotropic Trap Quench in $d=3$}

\begin{figure*}[htbp]
      \includegraphics[width=1\textwidth]{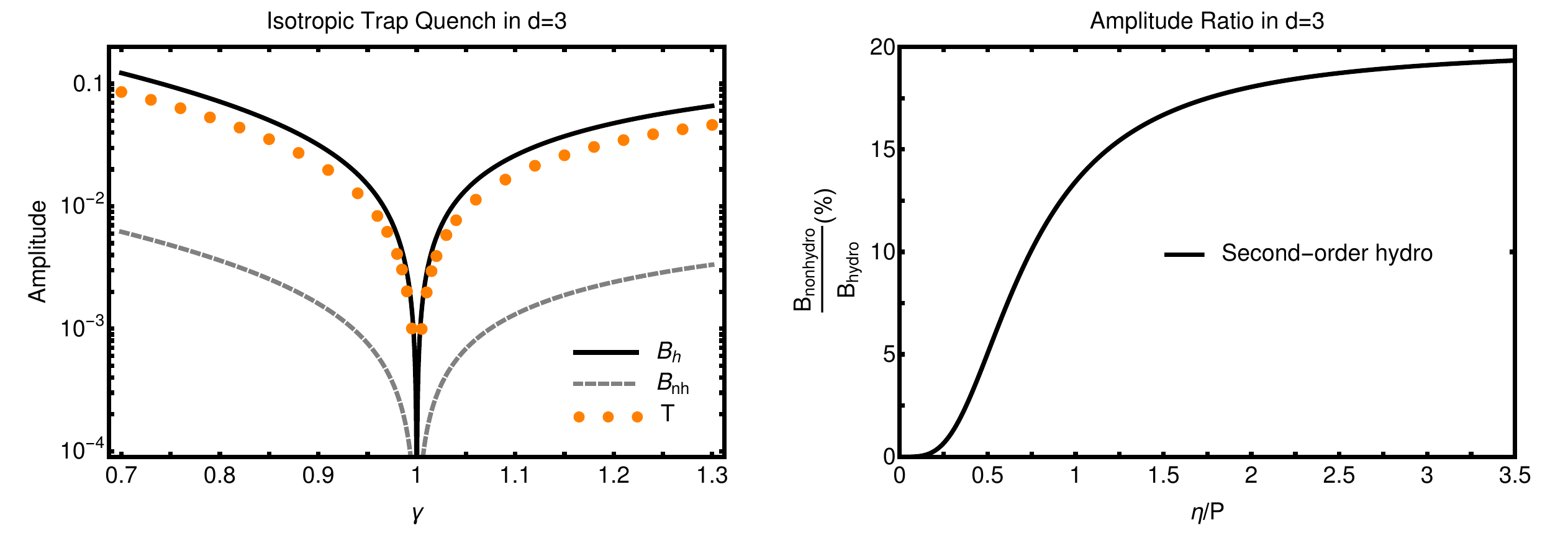}
     \caption{Left: Absolute value of the (dimensionless) hydrodynamic breathing (``B$_h$"),  non-hydrodynamic breathing (``B$_{nh}$") and temperature (``T") modes as a function the quench strength parameter $\gamma$ for an isotropic trap quench in d=3. Results shown are for $\frac{\eta}{P}=0.5$. Right: Ratio of the non-hydrodynamic to hydrodynamic breathing mode amplitudes as a function of $\eta/P$ for an isotropic trap quench in d=3 (result independent of $\gamma$). The maximal ratio of about $20\%$ is roughly consistent with the results of Ref.~\cite{BrewerPRL2015}, and suggests the possibility of experimental observation.}
      \label{fig:3diso}
\end{figure*}

For the case of an isotropic trap quench ($A_i/A_0=1,T_{\, init}=1$) in d=3, results for the (hydrodynamic and non-hydrodynamic) breathing mode and temperature mode amplitudes are shown in Fig.~\ref{fig:3diso}. Unlike the case of d=2, the three-dimensional geometry is capable of supporting a non-hydrodynamic breathing mode; furthermore, we find the temperature mode amplitude to differ from the (hydrodynamic) breathing mode amplitude. 

The right panel of Fig.~\ref{fig:3diso} shows the ratio of non-hydrodynamic to hydrodynamic breathing mode amplitudes, which reaches up to about $20\%$ for large enough $\eta/P$. It is interesting to note that the apparent saturation of this ratio at about $20\%$ is consistent with the amplitude ratio from Ref.~\cite{BrewerPRL2015}, extracted from experimental data in Ref.~\cite{KinastPRL2005}. (Note that in the experiment of Ref.~\cite{KinastPRL2005}, the gas was released from a symmetric trap, allowed to expand for a short period, and then recaptured in a symmetric trapping potential, which is a different protocol than the trap quench considered here. For this reason we do not attempt a direct comparison to mode amplitudes of Ref.~\cite{BrewerPRL2015}  in this case.)

\section{Conclusion}
\label{sec:conc}

We have utilized a second-order hydrodynamics framework in order to gain analytic insight into the collective response of Fermi gases in both two-dimensional (``pancake'') and three-dimensional (``cigar'') trap geometries.  Our results demonstrate a number of interesting features which may be reasonably expected to qualitatively describe experimental results. In some cases we even expect quantitative agreement, such as was the case for a Fermi gas undergoing an isotropic trap quench where we found our results to match those obtained from an exact quantum mechanical scaling solution \cite{Sergej2012}.

For instance, our analysis demonstrates that the damping rate of the volume conserving higher-harmonic modes is proportional to the shear viscosity times the harmonic mode number (i.e. the mode winding number $w$, see App.~\ref{App1}). Similar features have been predicted and experimentally observed in the context of relativistic heavy-ion collisions, cf. Refs.~\cite{Schenke2012,LaceyPRL2014}. While higher-order mode excitations likely result in weaker signal-to-noise ratio, our work suggests a potential experimental avenue towards a precision extraction of shear viscosity in cold Fermi gases by analogy with the relativistic heavy-ion results.

Our study also discusses the presence, damping rates, and expected mode amplitudes of non-hydrodynamic modes in trapped Fermi gases in detail. Recent studies on these non-hydrodynamic modes suggest they could provide information about the presence of quasi-particles in strongly coupled Fermi gases as well as exhibit deep connections between atomic physics and string theory \cite{BrewerPRL2015, Bantilan:2016qos}. The present results for the non-hydrodynamic mode amplitudes suggest that non-hydrodynamic modes should be accessible by state-of-the-art cold Fermi gas experiments.

There are a number of ways that the results presented here can be generalized and improved. For instance, we plan to study the case of collective modes for  uniform density and temperature equilibrium configurations (``Fermi gas in a box''), as well as anisotropic trap geometries in the future. Furthermore, the availability of fully numerical second-order hydrodynamic algorithms \cite{Bluhm2015,BrewerPRA2016} will allow relaxing the present assumptions of constant $\eta/P$, ideal equation of state, and small perturbations around equilibrium in order to aim for a fully quantitative exploration of the collective modes of trapped Fermi gases.

\acknowledgements{
This work was supported in part by the Department of Energy, DOE award No. DE-SC0008132. Publication of this article was funded by the University of Colorado Boulder Libraries Open Access Fund. PR would like to thank John Thomas for fruitful discussions as well as the organizers of the workshop on ``Non-Equilibrium Physics and Holography'' at Oxford University in July 2016 for providing a stimulating environment for many interesting discussions on this topic. WL would like to thank Jasmine Brewer and Roman Chapurin for useful comments regarding clarity of the manuscript.}

\appendix \section{Spatial Mode Structure for $d=2$}\label{App1}

In this appendix we collect the spatial mode structure for the low lying modes in $d=2$ (see Tab.~\ref{tab:spatialstruct2d}). It is interesting to note that, of the modes found, only the temperature mode and breathing mode are associated with a non-zero value of $\delta T$. This is because they are the only two modes which change the volume of the cloud, and hence can lead to heating and cooling of the gas. Also note that each irrotational volume conserving mode (here: dipole, quadrupole, hexapole, octupole, and decapole modes) has two independent realizations related by an appropriate coordinate transformation. For example, the quadrupole mode and tilted quadrupole mode are related by a rotation of $45^o$. In general, we can assign a mode the winding number $w$ of the associated velocity field $\mathbf{u}$ on a circle centered on the origin. This quantity merely counts the number of full rotations made when following a vector around the prescribed circle. For example, $w_{Dip}=0$, $w_{Quad}=1$, $w_{Hex}=2$,... so that the angle of coordinate rotations to get the second independent mode for a given mode is conveniently given by
	\begin{equation}
	\label{rotang}
		\Delta \phi = \frac{\pi}{2 (w+1)}.
	\end{equation}
Of course, any rotation through an angle in the range $\Delta \theta \in (0,\pi/(w+1))$ will produce an equally valid independent mode, but the angle in Eq.~\eqref{rotang} provides a uniform approach to finding an independent mode from one already found. This provides a connection for example to the approach in Ref.~\cite{BrewerPRL2015} where inequivalent polynomials under rotation up to quadrupole mode were considered in the second-order hydrodynamics framework used here.

\begin{table*}[h]
\begin{center}
\begin{tabular}{| c | c | c | c | c |}
  \hline
  \phantom{...} & $\delta\rho$ & $\delta u_x$ & $\delta u_y$ & $\delta T$\\
  \hline				
  Number (Zero Mode) & $1$ & $0$ & $0$ & $0$\\
  \hline			
    Temp. (Zero Mode) & $x^2+y^2-2$ & $0$ & $0$ & $2$\\
  \hline		
  Rotation (Zero Mode) & $0$ & $y$ & $-x$ & $0$\\
  \hline			
  Breathing (Monopole) & $x^2+y^2-2$ & $-i x \widetilde{\omega}_B $ & $- i y \widetilde{\omega}_B$ & $-2$\\
  \hline
  x-axis Sloshing (Dipole) & $x$ & $-i\widetilde{\omega}_S$ & 0 & 0 \\
  \hline
  y-axis Sloshing (Dipole) & $y$ & 0 & $-i\widetilde{\omega}_S$ & 0 \\
  \hline
  Quadrupole & $y^2-x^2$ & $i x \widetilde{\omega}_Q$ & $-i y \widetilde{\omega}_Q$ & 0\\
  \hline
  Tilted Quadrupole & $xy$ & $ -i \frac{y}{2} \widetilde{\omega}_Q$ & $-i 
 \frac{x}{2}\widetilde{\omega}_Q$ & 0\\
  \hline
  Hexapole & $y^3-3x^2y$ & $2ixy\widetilde{\omega}_H$  & $-i(y^2-x^2)\widetilde{\omega}_H$ & 0\\
    \hline
  T-Hexapole & $-\frac{x^3}{3}+xy^2$ & $-i\frac{(y^2-x^2)}{3}\widetilde{\omega}_H$  & $-i\frac{2xy}{3}\widetilde{\omega}_H$ & 0\\
  \hline
  Octupole & $x^4-6x^2y^2+y^4$ & $3i(-\frac{x^3}{3}+xy^2)\widetilde{\omega}_O$  & $-i(y^3-3x^2y)\widetilde{\omega}_O$ & 0\\
  \hline
    T-Octupole & $xy^3-x^3y$ & $i\frac{(3x^2y-y^3)}{4}\widetilde{\omega}_O$  & $i\frac{(x^3-3xy^2)}{4}\widetilde{\omega}_O$ & 0\\
  \hline
  Decapole & $y^5-10x^2y^3+5x^4y$ & $4i(xy^3-x^3y)\widetilde{\omega}_D$  & $-i(y^4-6x^2y^2+x^4)\widetilde{\omega}_D$ & 0\\
  \hline  
    T-Decapole & $\frac{x^5}{5}-2x^3y^2+xy^4$ & $-i\frac{(y^4-6x^2y^2+x^4)}{5}\widetilde{\omega}_D$  & $-4i\frac{(xy^3-x^3y)}{5}\widetilde{\omega}_D$ & 0\\
  \hline 
  Non-hydro Quad.& $y^2-x^2$ & $i x \widetilde{\omega}^{nh}_Q$ & $-i y \widetilde{\omega}^{nh}_Q$ & 0\\
    \hline  
  T-Non-hydro Quad.& $xy$ & $ -i \frac{y}{2} \widetilde{\omega}^{nh}_Q$ & $-i 
 \frac{x}{2}\widetilde{\omega}^{nh}_Q$ & 0\\
   \hline 
  Non-hydro Hex.& $y^3-3x^2y$ & $2ixy\widetilde{\omega}^{nh}_H$  & $-i(y^2-x^2)\widetilde{\omega}^{nh}_H$ & 0\\
     \hline 
  Non-hydro Oct.& $x^4-6x^2y^2+y^4$ & $3i(-\frac{x^3}{3}+xy^2)\widetilde{\omega}^{nh}_O$  & $-i(y^3-3x^2y)\widetilde{\omega}^{nh}_O$ & 0\\
  \hline 
  Non-hydro Dec.& $y^5-10x^2y^3+5x^4y$ & $4i(xy^3-x^3y)\widetilde{\omega}^{nh}_D$  & $-i(y^4-6x^2y^2+x^4)\widetilde{\omega}^{nh}_D$ & 0\\
  \hline
\end{tabular}
\end{center}
  \caption{Spatial structure of the various modes for $d=2$ expressed in terms of the normalized complex mode frequencies. Note that the tilted modes denoted Tilted- or T- for short in the table can be found by an appropriate rotation of coordinates.
}
	\label{tab:spatialstruct2d}
\end{table*}

\section{Spatial Mode Structure for $d=3$}\label{App2}

The spatial structure of modes in $d=3$ are given in Tab.~\ref{tab:spatialstruct3d}. Results are very similar to those for $d=2$ shown in App. \ref{App1}. The only differences are that for a harmonic trapping potential which is translationally invariant along one axis and isotropic along the other two in $d=3$, the breathing mode now couples to shear stresses. Hence there is now an associated non-hydrodynamic breathing mode as well as a difference in the corresponding temperature perturbation associated with the volume change of the cloud. Since there is no associated velocity field for the time independent temperature zero mode, this mode is associated with a vanishing stress tensor, and hence the mode structure is the same as in the $d=2$ case. All other modes also exhibit the same spatial and frequency structure.

\begin{table*}[h]
\begin{center}
\begin{tabular}{| c | c | c | c | c |}
  \hline
  \phantom{...} & $\delta\rho$ & $\delta u_x$ & $\delta u_y$ & $\delta T$\\
  \hline				
  Number (Zero Mode) & $1$ & $0$ & $0$ & $0$\\
  \hline			
    Temp. (Zero Mode) & $x^2+y^2-2$ & $0$ & $0$ & $2$\\
  \hline		
  Rotation (Zero Mode) & $0$ & $y$ & $-x$ & $0$\\
  \hline			
  Breathing (Monopole) & $x^2+y^2-2$ & $-i x \widetilde{\omega}_B $ & $- i y \widetilde{\omega}_B$ & $-\frac{4}{3}$\\
  \hline
  x-axis Sloshing (Dipole) & $x$ & $-i\widetilde{\omega}_S$ & 0 & 0 \\
  \hline
  y-axis Sloshing (Dipole) & $y$ & 0 & $-i\widetilde{\omega}_S$ & 0 \\
  \hline
  Quadrupole & $y^2-x^2$ & $i x \widetilde{\omega}_Q$ & $-i y \widetilde{\omega}_Q$ & 0\\
  \hline
  Tilted Quadrupole & $xy$ & $ -i \frac{y}{2} \widetilde{\omega}_Q$ & $-i 
 \frac{x}{2}\widetilde{\omega}_Q$ & 0\\
  \hline
  Hexapole & $y^3-3x^2y$ & $2ixy\widetilde{\omega}_H$  & $-i(y^2-x^2)\widetilde{\omega}_H$ & 0\\
    \hline
  T-Hexapole & $-\frac{x^3}{3}+xy^2$ & $-i\frac{(y^2-x^2)}{3}\widetilde{\omega}_H$  & $-i\frac{2xy}{3}\widetilde{\omega}_H$ & 0\\
  \hline
  Octupole & $x^4-6x^2y^2+y^4$ & $3i(-\frac{x^3}{3}+xy^2)\widetilde{\omega}_O$  & $-i(y^3-3x^2y)\widetilde{\omega}_O$ & 0\\
  \hline
    T-Octupole & $xy^3-x^3y$ & $i\frac{(3x^2y-y^3)}{4}\widetilde{\omega}_O$  & $i\frac{(x^3-3xy^2)}{4}\widetilde{\omega}_O$ & 0\\
  \hline
  Decapole & $y^5-10x^2y^3+5x^4y$ & $4i(xy^3-x^3y)\widetilde{\omega}_D$  & $-i(y^4-6x^2y^2+x^4)\widetilde{\omega}_D$ & 0\\
  \hline  
    T-Decapole & $\frac{x^5}{5}-2x^3y^2+xy^4$ & $-i\frac{(y^4-6x^2y^2+x^4)}{5}\widetilde{\omega}_D$  & $-4i\frac{(xy^3-x^3y)}{5}\widetilde{\omega}_D$ & 0\\
      \hline			
 Non-hydro Breath.& $x^2+y^2-2$ & $-i x \widetilde{\omega}^{nh}_B $ & $- i y \widetilde{\omega}^{nh}_B$ & $-\frac{4}{3}$\\
  \hline 
  Non-hydro Quad.& $y^2-x^2$ & $i x \widetilde{\omega}^{nh}_Q$ & $-i y \widetilde{\omega}^{nh}_Q$ & 0\\
    \hline  
  T-Non-hydro Quad.& $xy$ & $ -i \frac{y}{2} \widetilde{\omega}^{nh}_Q$ & $-i 
 \frac{x}{2}\widetilde{\omega}^{nh}_Q$ & 0\\
   \hline 
  Non-hydro Hex.& $y^3-3x^2y$ & $2ixy\widetilde{\omega}^{nh}_H$  & $-i(y^2-x^2)\widetilde{\omega}^{nh}_H$ & 0\\
     \hline 
  Non-hydro Oct.& $x^4-6x^2y^2+y^4$ & $3i(-\frac{x^3}{3}+xy^2)\widetilde{\omega}^{nh}_O$  & $-i(y^3-3x^2y)\widetilde{\omega}^{nh}_O$ & 0\\
  \hline 
  Non-hydro Dec.& $y^5-10x^2y^3+5x^4y$ & $4i(xy^3-x^3y)\widetilde{\omega}^{nh}_D$  & $-i(y^4-6x^2y^2+x^4)\widetilde{\omega}^{nh}_D$ & 0\\
  \hline
\end{tabular}
\end{center}
  \caption{Spatial structure of the various modes for $d=3$ expressed in terms of the normalized complex mode frequencies. Note that the tilted modes denoted Tilted- or T- for short in the table can be found by an appropriate rotation of coordinates.
}
	\label{tab:spatialstruct3d}
\end{table*}

\section{Details of Mode Amplitude Calculation}\label{App3}

Given generic initial conditions on $\rho$, $\mathbf{u}$, $T$, and $\pi_{ij}$, one can derive a system of equations for complex mode amplitudes $(a_n + i b_n)$ of mode $n$ (e.g. $n=$``number'', ``temperature'', ``breathing'', etc.) by performing the following projections onto a mode $m$:
\begin{align}
	\label{proj1}&\int_{\mathbb{R}^2} d^2\mathbf{x} \bigg[\rho_{\,init}(\mathbf{x}) -\rho_0(\mathbf{x}) \bigg]\delta \rho_m(\mathbf{x})  =\\
  \nonumber &\int_{\mathbb{R}^2} d^2\mathbf{x}   \rho_0(\mathbf{x}) \sum_{modes\, n}\mathcal{R}e\bigg[(a_n + i b_n) e^{-i \omega_n t}\delta \rho_n(\mathbf{x})\bigg]\delta \rho_m(\mathbf{x}),\\
	\label{proj2} &\int_{\mathbb{R}^2} d^2\mathbf{x} \bigg[u_{i \phantom{.} \,init}(\mathbf{x}) \bigg]\rho_0(\mathbf{x}) \delta u_{im}(\mathbf{x}) =\\
\nonumber&\int_{\mathbb{R}^2} d^2\mathbf{x} \sum_{modes\, n} \mathcal{R}e\bigg[(a_n + i b_n) e^{-i \omega_n t}\delta\mathbf{u}_n(\mathbf{x})\bigg]\rho_0(\mathbf{x}) \delta u_{im}(\mathbf{x}),\\
	\label{proj3}&\int_{\mathbb{R}^2} d^2\mathbf{x} \bigg[T_{\,init}-T_{0}\bigg] \rho_0(\mathbf{x}) \delta T_{m} =\\
  \nonumber&\int_{\mathbb{R}^2} d^2\mathbf{x} \sum_{modes\, n} \mathcal{R}e\bigg[(a_n + i b_n) e^{-i \omega_n t} \delta T_{n}\bigg]\rho_0(\mathbf{x})\delta T_{m},
	\end{align}
	\begin{align}
	\label{proj4}&\int_{\mathbb{R}^2} d^2\mathbf{x} \bigg[\pi_{ij \phantom{.} \,init}(\mathbf{x}) \bigg] \delta \pi_{ij\phantom{.}m}(\mathbf{x}) =\\
\nonumber &\int_{\mathbb{R}^2} d^2\mathbf{x} \sum_{modes\, n}\mathcal{R}e\bigg[(a_n + i b_n) e^{-i \omega_n t}\delta \pi_{ij\phantom{.}n}(\mathbf{x})\bigg] \delta \pi_{ij\phantom{.}m}(\mathbf{x}),\,.
	\end{align}
It should be noted that as can be seen from Tabs.\ref{tab:spatialstruct2d} and \ref{tab:spatialstruct3d}, not all of the individual perturbations are orthogonal (the full mode structures are, however, independent). For example, while $\delta\rho_{Temp} = \delta\rho_{B}$, it is not possible to construct the full mode structure $\delta_{Temp}=\{\delta \rho_{Temp},\delta \mathbf{u}_{Temp}, \delta T_{\phantom{.}Temp} \}$ as a linear combination of the full mode structure of the other modes. We note that as a result, care should be used when obtaining the system of equations for the amplitudes give by evaluating Eqs.~\eqref{proj1}-\eqref{proj4} not to miss contributions from all the important modes. We also point out that in general our process for finding mode structure, it is found that mode frequencies come in pairs, one with positive and the other with negative real part. However, in allowing for a complex amplitude and taking the real part in Eqs.~\eqref{proj1}-\eqref{proj4} we need only consider modes to have positive real part of the frequency.

Let us consider a generic isotropic trap quench in $d=2$ for multiple initial conditions in order to demonstrate the role of the temperature and number modes. In this case, the amplitudes take on a fairly simple form 
\begin{align}
	\label{numamp}
	&a_N=\frac{A_i}{A_0}-1,\\
	&a_{T}=\frac{(1-\gamma)\frac{A_i}{A_0}-(T_{\,init}-1)\gamma}{4 \gamma},\\
	&a_{B}=\frac{(1-\gamma)\frac{A_i}{A_0}-(1-T_{\,init})\gamma}{4 \gamma},\\
	\label{bamp}	
	&b_{B}=0.
\end{align}
Note that the above expressions contain both phase and magnitude information, as they may be negative. We will plot phase and amplitude separately below. Additionally, we see from Eqs.~\eqref{numamp}-\eqref{bamp} several features which should be expected. To explore this, we will break up our analysis into several cases. \\

\noindent \textbf{Case 1: $\mathbf{A_i/A_0=1}$, $\mathbf{T_{\,init}=1}$}

This is the case discussed in the main text (see Sec.~\ref{2diso}). \\

\noindent \textbf{Case 2: $\mathbf{A_i/A_0\neq1}$, $\mathbf{T_{\,init}=1}$}

\begin{figure*}[htbp]
      \includegraphics[width=\textwidth]{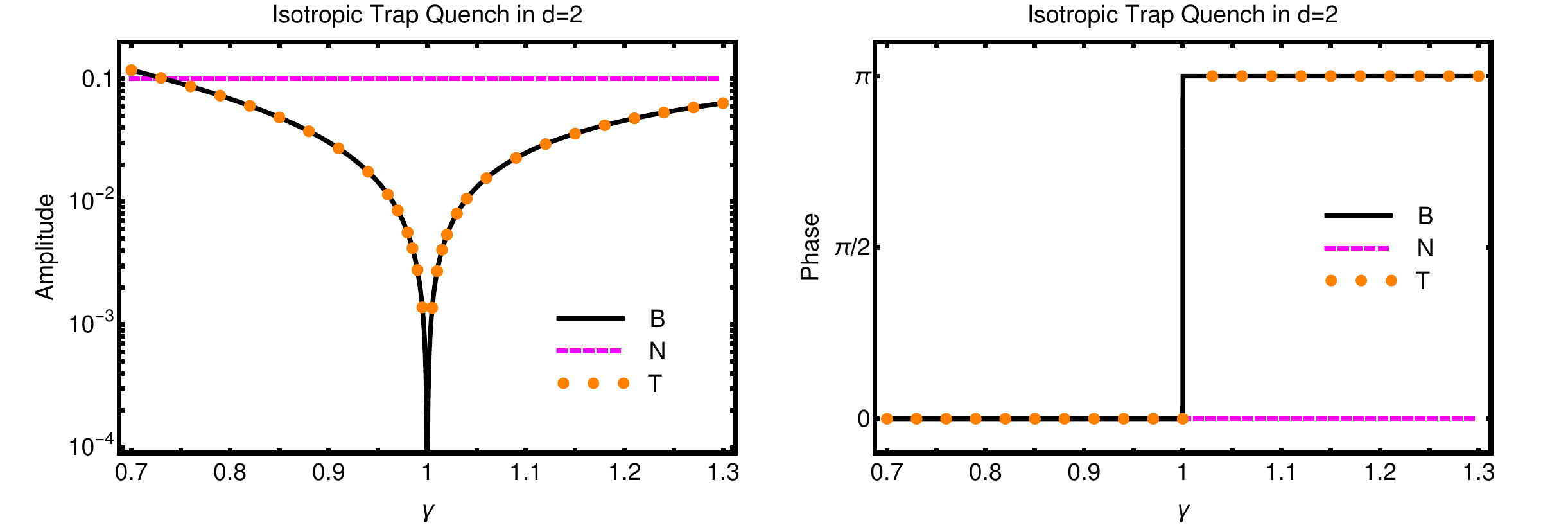}
      \caption{Left: Absolute value of the (dimensionless) number ('N'), temperature ('T') and breathing ('B') mode amplitudes as a function of the quench strength parameter $\gamma$ for an isotropic quench in d=2 assuming $A_i/A_0=1.1>1$ and $T_{\,init}=1$. Right: Phase of mode amplitude. Note the amplitude and phase of the breathing and temperature modes are identical.}
      \label{more}
\end{figure*}

For this case we see from Eqs.~\eqref{numamp}-\eqref{bamp} that the ratio $A_i/A_0$ gives rise to a non-zero amplitude of the number mode, but leaves the location of the zero of the other two modes at $\gamma=1$. This should be expected since this merely means that at $\gamma=1$ we have more ($A_i/A_0>1$) or less ($A_i/A_0<1$) atoms in the trap than what was assumed in the equilibrium we expanded about. This should make the role of the number mode more clear and is demonstrated for the case $A_i/A_0>$ and $T_{init}=1$ in Fig.~\ref{more}. Particularly, it is only important if for some reason we chose to expand our dynamics about an equilibrium with a different number of particles than given by our initial conditions.

\noindent \textbf{Case 3: $\mathbf{A_i/A_0=1}$, $\mathbf{T_{\,init}\neq1}$}

\begin{figure*}[htbp]
      \includegraphics[width=\textwidth]{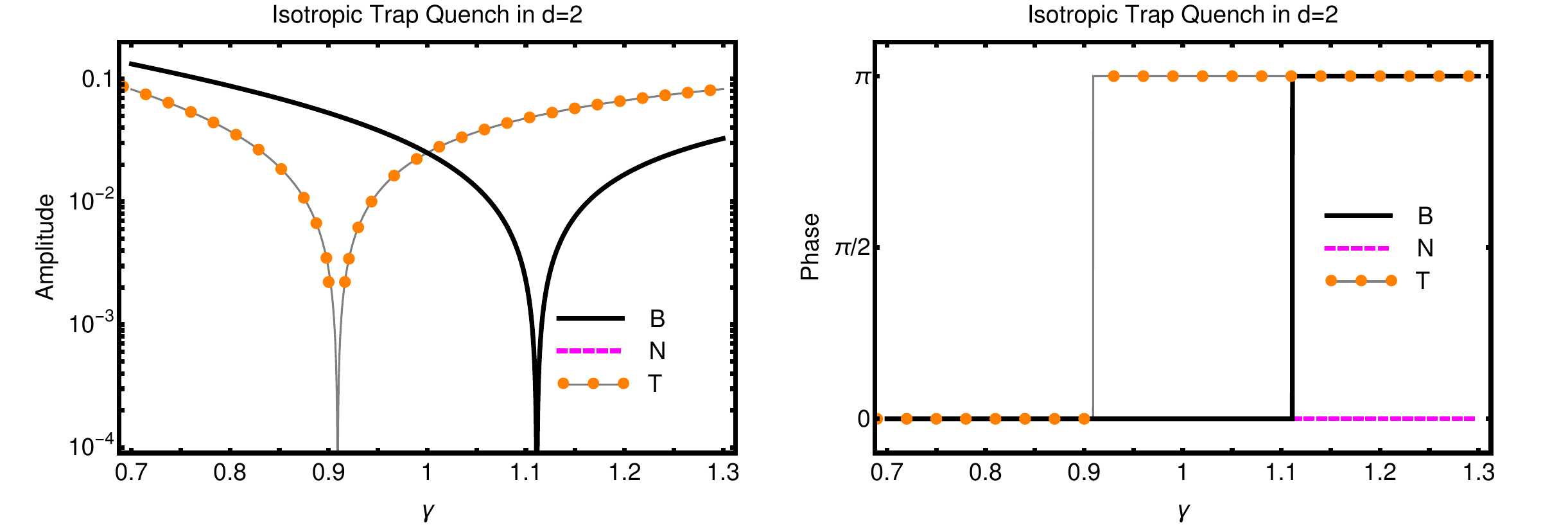}
      \caption{Left: Absolute value of the (dimensionless) temperature ('T') and breathing ('B') mode amplitudes as a function of the quench strength parameter $\gamma$ for an isotropic quench in d=2 assuming $A_i/A_0=1$ and $T_{\,init}=0.9$. Right: Phase of mode amplitude.}
      \label{lowerT}
\end{figure*}

For this case we see from Eqs.~\eqref{numamp}-\eqref{bamp} that the number mode is not excited, while the value of $T_{\,init}$ alters the location of the zero for the temperature and breathing modes. This should be expected since at $\gamma=1$ the breathing mode is excited through the temperature difference. Fig.~\ref{lowerT} shows the case where $T_{\,init}<1$, demonstrating that the phase of the breathing mode vanishes at $\gamma=1$ while the magnitude is non-zero. A positive amplitude at $\gamma=1$  is expected since the temperature is below its equilibrium value for the given cloud radius so the cloud will reduce its size to try to reach equilibrium.


\newpage
\bibliography{coldmodes2}

\end{document}